 \documentclass[cameraready]{Interspeech}
 \usepackage{cite}

\title{Comparative Reasoning: Making an Audio Language Model Better at Comparing Emotions}

\author[affiliation={1,2}, orcid=0000-0003-2686-6278, equalcontribution]{Abinay Reddy}{Naini}
\author[affiliation={1}, orcid=0009-0007-4412-9999, equalcontribution, ]{Jaeyeon}{Kim}
\author[affiliation={3}, orcid=0000-0003-2879-8811]{Chao-Han Huck}{Yang}
\author[affiliation={1}, orcid=0000-0002-5970-8631]{Shinji}{Watanabe}
\author[affiliation={1}, orcid=0000-0002-4075-4072, correspondingauthor]{Carlos}{Busso}

\address{
    $^1$ Language Technologies Institute, Carnegie Mellon University, Pittsburgh, PA, 15213, US\\
    $^2$ The University of Texas at Dallas, Richardson TX 75080, USA;  
    $^3$ NVIDIA
}

\email{abinayreddy.naini@utdallas.edu, jaeyeon2@andrew.cmu.edu, hucky@nvidia.com, shinjiw@ieee.org, busso@cmu.edu}


\usepackage{comment}
\usepackage{booktabs}
\usepackage{tabularx}
\usepackage{color}
\usepackage{bm}
\usepackage{colortbl}
\usepackage{amsmath}
\usepackage{amssymb}
\usepackage[most]{tcolorbox}    
\tcbuselibrary{skins, breakable} 
\usepackage{multicol}           
\usepackage{xcolor}             
\usepackage{CJKutf8} 

\newtcolorbox{promptbox}[2][]{%
    enhanced,
    colback=gray!5!white,
    colframe=gray!60!black,
    fonttitle=\footnotesize,
    fontupper=\footnotesize,
    title={#2},
    attach boxed title to top left={yshift=-3mm, xshift=1mm},
    boxed title style={
        colback=gray!60!black, 
        arc=1mm, 
    },
    arc=3mm, 
    boxrule=0.5pt,
    top=4mm, 
    bottom=2mm,
    left=2mm,
    right=2mm,
    width=\linewidth, 
    boxsep=1pt,
    #1
}

\begin{document}

\maketitle

\begin{abstract} 
\emph{Large audio-language models} (LALMs) can reason about audio, yet it remains unclear whether they can perform comparative judgments between two speech signals along emotional, environmental, linguistic, prosodic, and interpersonal dimensions. We study this question in the context of \emph{speech emotion recognition} (SER), where the model determines which utterance exhibits higher arousal, valence, or dominance. We introduce a reasoning-guided ordinal SER framework that conditions an LALM on paired speech inputs. The model is trained using reasoning traces generated from both semantic audio descriptions and acoustic evidence derived from GeMAPS features, enabling interpretable comparative decisions. Beyond direct supervision, we also employ direct preference optimization to encourage stronger separation for emotional differences. Experiments show that the proposed framework improves preference prediction while requiring only 5\% of the training data used by conventional ordinal SER systems.
\end{abstract}

\section{Introduction}
\label{sec:introduction}
Recent progress in \emph{large audio-language models} (LALMs) has substantially advanced audio understanding across a wide range of domains, including speech, environmental sounds, music, and complex acoustic scenes \cite{Chu2024QwenAudio, tang2023salmonn, zhang2023speechgpt, cheng2024emotion, af3}. Recent studies have further enhanced these models with reasoning capabilities through \emph{chain-of-thought} (CoT) supervision \cite{wei2022chain, audio_reasoner} and reinforcement learning–based post-training \cite{ouyang2022training, rafailov2023direct, deepseek_math, omni_r1, audio_thinker}, enabling intermediate inference reasoning steps before prediction to further solve complex questions. Despite these advances, comparative reasoning across multiple audio inputs remains largely underexplored, and it is unclear whether these models can provide interpretable explanations for why one audio signal should be preferred over another. For example, a model may need to determine which utterance is more emotionally intense, louder, noisier, or exhibits greater pitch variation, yet most existing LALMs are primarily designed for single-audio inference with limited support for multi-audio comparison \cite{chu2024qwen2audiotechnicalreport, adiff, af3}. Correspondingly, emerging benchmarks that require comparative reasoning over multiple audio signals report that current LALMs exhibit relatively weak performance in comparative setups \cite{mmau-pro, wow_bench, adiff, mae}.

To systematically investigate comparative reasoning in LALMs, we focus on preference-based \emph{speech emotion recognition} (SER) as a targeted evaluation setting. 
While SER has been traditionally formulated as a classification or regression task \cite{Leem_2022,Leem_2024, Lotfian_2017}, emotional perception offers a natural testbed for comparative reasoning because emotions are perceived relatively rather than absolutely. Psychological studies show that humans struggle to assign consistent absolute emotional scores but demonstrate significantly higher agreement when comparing stimuli pairwise  \cite{Helson_1964}. This observation has motivated preference learning approaches in SER, where models determine whether one utterance expresses a stronger emotional attribute (e.g., higher arousal or more positive valence) or a stronger emotional category (e.g., one utterance is happier than the other) \cite{Cao_2012, Abdelaziz_2017, Yannakakis_2017,Yannakakis_2021}. These comparative annotations have been shown to better align with human judgments and reduce the impact of annotation variability \cite{Parthasarathy_2017,Lei_2023, Naini_2023}. Preference-based emotion recognition also inherently requires comparative reasoning over multiple cues. Determining relative emotional intensity involves evaluating acoustic properties such as pitch level, loudness variation, and speaking rate, together with semantic and paralinguistic information conveyed by speech. The model must, therefore, interpret evidence from two signals jointly rather than analyze each utterance in isolation. In addition to predicting which utterance is preferred, such comparisons also require identifying perceptually relevant cues that justify the decision. As such, emotion comparison provides a controlled yet cognitively grounded framework for evaluating whether LALMs can perform meaningful and interpretable cross-audio reasoning.

How to effectively adapt LALMs, despite their strong performance in other downstream tasks, to this preference-based setting remains an open problem. Existing emotion preference learning approaches learn comparative relationships solely from annotation-derived labels. \emph{Self-supervised learning} (SSL) based preference frameworks \cite{Naini_2024} estimate whether one sample expresses higher arousal or more positive valence than another, but they do not model how such comparisons are internally constructed. As a result, comparative decisions are learned implicitly from representations rather than through explicit reasoning over perceptually relevant cues. 

\begin{figure*}[t]
\centering
\includegraphics[width=0.85\textwidth]{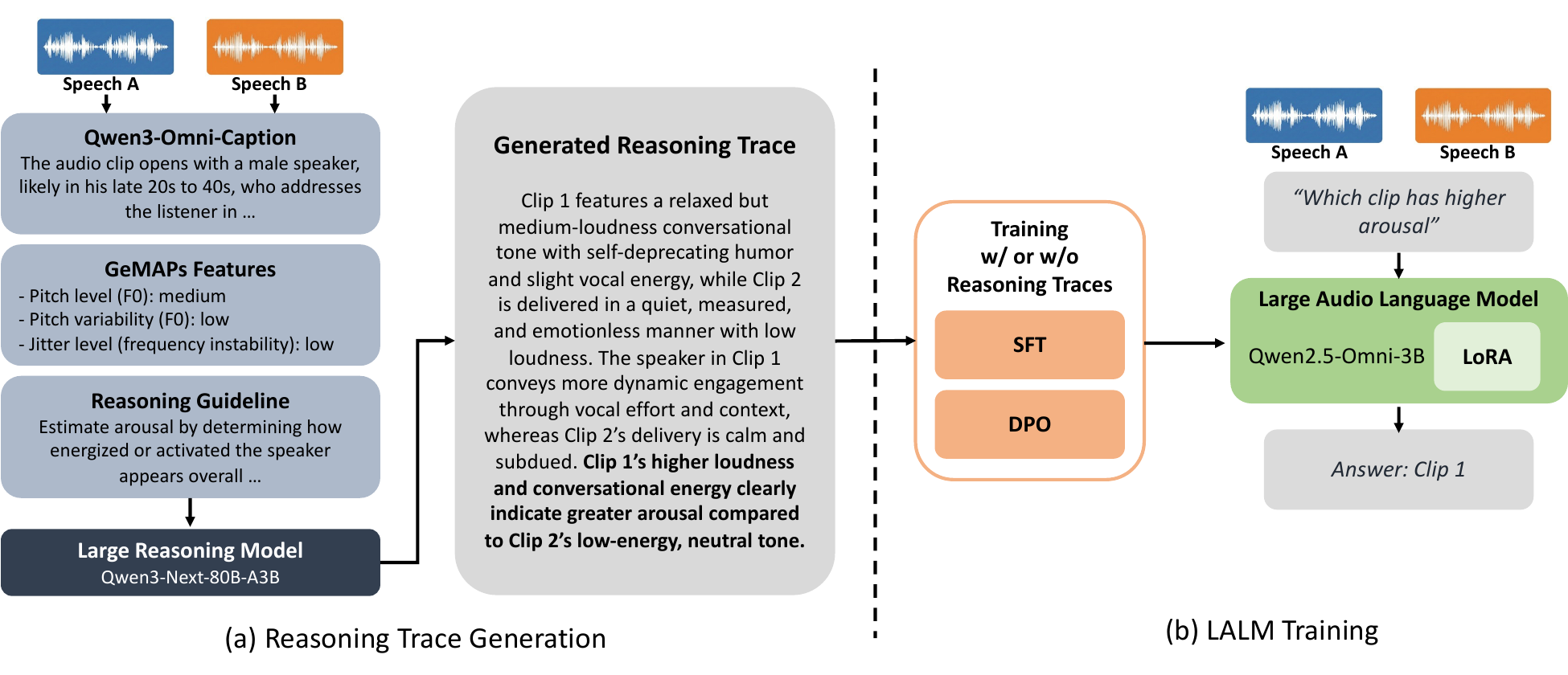}
\vspace{-0.4cm}
\caption{
Overview of the proposed reasoning-guided ordinal speech emotion recognition framework.
(a) Reasoning trace generation: acoustic GeMAPS features and audio descriptions are provided to a large reasoning model to produce a comparative reasoning trace.
(b) LALM training: the audio-language model receives two speech inputs and predicts the relative emotional attribute using supervised fine-tuning (SFT) and direct preference optimization (DPO).
}
\label{Block_pref}
\vspace{-2mm}
\end{figure*}

In this work, we study how LALMs can be adapted for interpretable comparative reasoning in ordinal speech emotion recognition. To guide these comparisons, we construct comparative reasoning traces that combine semantic audio descriptions with GeMAPS acoustic features  \cite{Eyben_2016}, providing rich perceptual grounding for the model's decisions. We explore multiple training paradigms, including \emph{supervised fine-tuning} (SFT) and \emph{direct preference optimization} (DPO) with and without reasoning traces. 
Experimental results show that properly adapted LALMs demonstrate strong potential for emotion preference learning, and that incorporating intermediate reasoning improves interpretability of model decisions.

\section{Related Work}

Preference learning has been explored in SER well before the emergence of audio-language models. Early studies showed that humans are more consistent when comparing emotional intensities than when assigning absolute ratings, motivating ordinal supervision for emotion perception \cite{Yannakakis_2017,Parthasarathy_2018_2,Yannakakis_2021,Yannakakis_2015}. Building on this observation, learning-to-rank formulations were introduced for categorical emotions, where models were trained to establish preferences among emotional classes \cite{Cao_2012,Lotfian_2016_2,Cao_2015}. Subsequent work focused on constructing reliable pairwise labels from multi-annotator emotional attribute scores and modeling comparative relationships using ranking-based formulations \cite{Martinez_2014,Parthasarathy_2016_2,Naini_2023_3,Naini_2025_3}. These approaches include agreement-aware labeling, trend-based comparisons across annotators, and other strategies for deriving stable ordinal supervision from subjective annotations \cite{Lotfian_2016_2,Parthasarathy_2018_2,Naini_2023_2}. Other ordinal formulations model emotion ordering directly using ordinal classification losses such as CORAL \cite{Han_2020}.

In parallel, recent work has begun incorporating reasoning and explainability into speech emotion modeling. Explainable speech-language model frameworks use reasoning traces to justify emotion predictions \cite{Su2025ReasoningBeyondMajorityVote}. Other approaches refine empathetic responses or optimize evidence-grounded emotional reasoning through reinforcement learning and agentic decoding \cite{Chen2026RELLM,Sun_2026_3,Wang2026EmotionThinker}. These studies demonstrate that reasoning can improve interpretability and robustness of emotion modeling by explaining predictions for individual speech samples. However, they focus on predicting or explaining emotions for individual utterances and do not address comparative reasoning between multiple speech samples. A small number of studies have examined preference prediction with language models. EmoPrefer \cite{Lian2025EmoPrefer} evaluates whether multimodal LLMs can select preferred emotional descriptions for multimedia content. This setting differs from ordinal SER, where preferences are defined over the relative emotional attributes expressed in speech signals rather than over textual emotion descriptions. To the best of our knowledge, the reasoning capability of LALMs has not been investigated within an ordinal SER framework that learns from pairwise emotional comparisons or explicitly models the strength of emotional preferences.

\section{Methodology}

Figure~\ref{Block_pref} presents an overview of the proposed reasoning-guided ordinal SER framework. The goal of the system is to determine the relative emotional attribute level between two speech samples. Given two utterances, the model predicts which sample exhibits higher arousal, valence, or dominance. Instead of estimating absolute attribute scores, the task is formulated as a comparative decision problem.

\vspace{-2mm}
\subsection{Framework Overview}
Let $x_A$ and $x_B$ denote two speech clips. A prompt is provided to the LALM asking which sample exhibits higher emotional intensity for a target attribute. 
For both zero-shot and trained settings, the prompt is structured to (1) explicitly specify the input order of the speech samples (Clip 1 and Clip 2), and (2) provide a definition of the target emotional dimension (e.g., characteristics of high versus low arousal), enabling the LALM to correctly interpret the comparative task. We denote this prompt as $p$.
The LALM then produces one of two possible responses indicating whether $x_A$ or $x_B$ has a higher attribute level. Let $y^+$ denote the correct response corresponding to the clip with higher emotional intensity, and $y^-$ denote the incorrect response. Given $(x_A, x_B, p)$ as input to the LALM, the task is to predict $y^+$.

\subsection{SFT and DPO with Labels}
As an initial setup, we first train the model directly using supervision labels. For SFT, given $(x_A, x_B, p)$, the model is trained to directly predict $y^+$. 
We additionally explore DPO, as its preference-based formulation may better distinguish characteristics between correct and incorrect responses. Given $(x_A, x_B, p)$, we construct preference pairs $(y^{+}) \succ (y^{-})$ and apply DPO to these pairs. Following prior work~\cite{pang2024iterative}, we further incorporate an SFT loss with a weight of 1.0 during DPO training.

\subsection{Comparative Reasoning Trace for Ordinal SER}
While the above SFT and DPO training enable the model to make direct comparative predictions, they do not explicitly guide how the comparison should be performed. Human listeners typically compare speech samples by interpreting acoustic cues such as pitch, loudness, and vocal stability. Therefore, we introduce a reasoning-guided strategy that provides structured perceptual evidence to the model prior to decision making.

Fig.~\ref{Block_pref}(a) illustrates the proposed reasoning framework. We first employ Qwen3-Omni-Captioner~\cite{xu2025qwen3omnitechnicalreport} to obtain detailed descriptions of the given speech samples. However, while these captions primarily capture semantic interpretation, they often lack fine-grained acoustic details and may introduce hallucinated content. 

In order to compensate for this limitation, we additionally extract 18 GeMAPS \emph{low-level descriptors} (LLDs) from each speech clips. For each LLD, both the mean and standard deviation are computed, resulting in a 36-dimensional acoustic representation. To produce interpretable descriptions from these features, feature values are normalized across the training dataset and discretized into qualitative levels. Each acoustic characteristic is mapped into descriptive categories such as \textit{low}, \textit{medium}, or \textit{high}, capturing perceptually meaningful properties including pitch level, pitch variability, loudness level, vocal stability, roughness, and spectral brightness. 

The semantic captions and acoustic feature descriptions for both audio samples are then provided to a large reasoning model, together with guidelines specifying which attributes should be considered when comparing the target emotional dimension. The reasoning model generates a structured reasoning trace $r^+$ that summarizes salient characteristics, performs comparative analysis, and produces the final decision $y$. We additionally constrain the reasoning trace to be concise, \textit{i.e.}, fewer than five sentences, as preliminary experiments showed that longer reasoning traces are more prone to hallucination and may lead to degraded performance.
The generated reasoning is verified by comparing the predicted decision $y$ with the ground-truth label $y^+$. If the reasoning does not lead to the correct answer, we regenerate the reasoning trace by additionally conditioning on the correct label $y^+$ together with the descriptions
Given $(x_A, x_B, p)$, under SFT training, the model is trained to generate both the reasoning trace and the final answer, \textit{i.e.}, $(r^+, y^+)$, instead of predicting only the label $y^+$.

Additionally, to utilize reasoning traces in DPO training, we prompt a large reasoning model with the audio descriptions and an incorrect answer $y^-$ to generate a corresponding reasoning trace $r^-$ that leads to the wrong decision. The model is not informed that the provided answer is incorrect, allowing it to generate a plausible reasoning trace that justifies the answer. We then construct preference pairs $(r^+, y^+) \succ (r^-, y^-)$ that enforce both correct reasoning and answers over incorrect reasoning and answers, and optimize the model using DPO.

\section{Experiments}

\subsection{Datasets and Label Preparation}
\label{ssec:data}

We conduct experiments primarily on the MSP-Podcast v2.0 corpus \cite{Busso2025MSPPodcast}, which contains approximately 409 hours of speech annotated for the emotional attributes (arousal, valence, and dominance), primary and secondary categorical emotions. The dataset is partitioned into training, development, and test sets with 169,190, 34,399, and 46,294 speech segments, respectively. Each segment is annotated by multiple raters using a 1–7 Likert scale. To evaluate cross-domain generalization, we additionally use the BIIC-Podcast corpus \cite{Upadhyay_2023_2}, which contains podcast recordings in Mandarin, and the WHiSER corpus \cite{Naini_2024_2}, consisting of 5,427 speech segments extracted from President Nixon's Oval Office recordings between 1971 and 1973 \cite{Nixons_tapes_1972}. Both corpora have similar emotional annotations as the MSP-Podcast corpus.

For ordinal learning, we construct preference pairs from the MSP-Podcast training set using approximately 5\% of the available utterances. Two samples $u_1$ and $u_2$ form a valid pair if the absolute difference between their consensus attribute scores satisfies $|m_{u_1}-m_{u_2}| > 1$ on the 1–7 scale. From this subset, we create a training set of 10k pairs for each emotional attribute (arousal, valence, and dominance). For evaluation, 3k pairs are sampled from the MSP-Podcast development set and 3k from the MSP-Podcast test set. We also construct additional test sets containing 3k pairs from the WHiSER and BIIC corpora to evaluate cross-domain generalization.

\vspace{-0.1cm}
\subsection{Experimental Setup and Metrics}

We use Qwen2.5-Omni-3B \cite{Chu2024QwenAudio} as the backbone LALM for all experiments. Reasoning traces are generated using Qwen3-Next-80B \cite{qwen_next80b_technical_report}. For parameter-efficient adaptation, we apply \emph{low-rank adaptation} (LoRA) \cite{hu_2021_lora} with rank $r=64$ and scaling factor $\alpha=64$ to all linear layers during SFT and DPO alignment. Performance is evaluated using attribute-specific preference accuracy for arousal, valence, and dominance (i.e., the percentage of pairs were the preference is properly recognized). We also report the average attribute preference accuracy computed across the three emotional dimensions. All evaluations are conducted on held-out preference pairs constructed as described in Section~\ref{ssec:data}.\\
\noindent\underline{SSL-based Ranking Baselines:}
We compare the proposed approach with conventional self-supervised speech representations trained using ranking-based preference learning. Specifically, we use WavLM \cite{Chen_2022} and HuBERT \cite{Hsu_2021} features combined with RankNet \cite{Burges_2005}, a pairwise learning-to-rank objective that predicts the probability that one sample should be preferred over another. We also include RankList \cite{Naini_2026}, a listwise preference learning framework that extends RankNet to jointly model ordinality across list of samples. These baselines are trained using 240k speech sample pairs for each emotional attribute.

\vspace{-0.1cm}
\subsection{Main Results}
\begin{table}[t]
\centering
\caption{Preference accuracy (\%) on MSP-Podcast test pairs. SFT/DPO-CoT denotes the model variants trained with reasoning traces.}
\vspace{-2mm}
\label{tab:main_results}
\resizebox{\linewidth}{!}{
\begin{tabular}{lcccc}
\hline
Model & Arousal & Valence & Dominance & Avg \\
\hline
WavLM + RankNet & 0.792 & 0.806 & 0.753 & 0.784 \\
HuBERT + RankNet & 0.781 & 0.773 & 0.742 & 0.765 \\
RankList \cite{Naini_2026}& 0.808 & 0.813 & 0.767 & 0.796 \\
\hline
Qwen2.5-Omni-3B & 0.658 & 0.707 & 0.547 & 0.637 \\
 + SFT & 0.881 & 0.878 & 0.867 & 0.875 \\
+ SFT-CoT & 0.855 & 0.865 & 0.846 & 0.855  \\
\hline
+ DPO & 0.885 & 0.888 & 0.863 & 0.879 \\
+ DPO-CoT & \textbf{0.887} & \textbf{0.890} & \textbf{0.867} & \textbf{0.881} \\
\hline
\end{tabular}
}
\end{table}

Table~\ref{tab:main_results} compares the proposed LALM framework with SSL-based ranking baselines. The zero-shot Qwen2.5-Omni-3B model performs poorly on ordinal comparison, indicating that pretraining alone does not provide reliable cross-audio emotional judgment. After SFT on the audio pairs, performance increases substantially and surpasses all SSL ranking baselines despite using far fewer training pairs (10k vs.\ 240k). These results suggest that LALM possess latent comparative capabilities when properly adapted to the task, and that LALMs can learn ordinal perception with significantly higher data efficiency than conventional representation-learning approaches. Notably, the improvements are particularly strong for the dominance attribute, which has historically been one of the most challenging emotional dimensions to model reliably in SER \cite{Wagner_2023, Naini_2024}.

Preference optimization further improves performance. The DPO consistently improves over SFT, suggesting that learning from correct versus incorrect decisions better matches the comparative nature of the task. 
Notably, while models trained with reasoning traces perform slightly worse than label-only models under SFT, they achieve better results after DPO training. This result indicates that explicitly distinguishing between correct reasoning traces $r^+$ and incorrect reasoning traces $r^-$ helps the model better utilize its reasoning capability and reduces hallucinated reasoning. Furthermore, models trained with reasoning traces provide improved interpretability and reliability because they explicitly explain their decisions, as shown in Fig.~\ref{fig:qualitative}. The results demonstrate that LALMs can reliably compare emotional attributes when properly adapted. 






\begin{table}[t]
\centering
\caption{Cross-domain preference accuracy (\%) on BIIC and WHiSER test pairs. SFT/DPO-CoT denotes the model variants trained with reasoning traces.}
\vspace{-2mm}
\label{tab:cross_domain}
\begin{tabular}{lcc}
\hline
Model & BIIC Avg & WHiSER Avg \\
\hline
WavLM + RankNet & 0.721 & 0.764 \\
HuBERT + RankNet & 0.710 & 0.758 \\
RankList \cite{Naini_2026}& 0.737 & 0.772 \\
\hline
Qwen2.5-Omni-3B & 0.572 & 0.677 \\
 + SFT & 0.760 & 0.898 \\
 + SFT-CoT & 0.741 & 0.854 \\
 \hline
+ DPO & 0.757 & \textbf{0.911} \\
+ DPO-CoT & \textbf{0.770} & 0.909 \\
\hline
\end{tabular}
\end{table}

\begin{table}[t]
\centering
\caption{Cross-emotion preference accuracy (\%) on MSP-Podcast test pairs. Each variants are trained on Arousal only. SFT/DPO-CoT denotes the model variants trained with reasoning traces.}
\vspace{-2mm}
\label{tab:cross_emotion}
\resizebox{\linewidth}{!}{
\begin{tabular}{lcccc}
\hline
Model & Arousal & Valence & Dominance & Avg \\
\hline
Qwen2.5-Omni-3B & 0.658 & \textbf{0.707} & 0.547 & 0.637 \\
+ SFT &  0.880 & 0.484 & 0.863 & 0.743 \\
+ SFT-CoT & 0.851 & 0.542 & 0.844 & 0.746  \\
\hline
+ DPO & 0.880 & 0.514 & 0.862 & 0.752 \\
+ DPO-CoT & \textbf{0.881} & 0.607 & \textbf{0.868} & \textbf{0.785} \\
\hline
\end{tabular}
}
\end{table}

\subsection{Cross-Domain Generalization}

We conduct cross-dataset and cross-emotion generalization experiments. For cross-dataset evaluation, models trained on the MSP-Podcast corpus are directly evaluated on the Whisper corpus and BIIC-Podcast corpus without additional training. For cross-emotion generalization, models are trained using only arousal preference pairs from the MSP-Podcast corpus and evaluated on other emotional dimensions.

Cross-dataset results are shown in Table~\ref{tab:cross_domain}. Overall, LALM-based models generalize better than SSL-based baselines and are more robust to dataset shifts. DPO-based training consistently performs better than SFT, suggesting that preference optimization improves robustness under domain transfer. Furthermore, DPO-CoT consistently outperforms SFT-CoT, suggesting that explicitly contrasting correct and incorrect reasoning traces helps the model learn more robust comparison strategies that generalize better across datasets.

Cross-emotion results are presented in Table~\ref{tab:cross_emotion}. Reasoning-based models generally outperform label-only models, showing noticeably less performance degradation on the valence dimension. This can be attributed to the model’s ability to reason about acoustic characteristics based on the definitions of emotional attributes (e.g., high vs. low valence) explicitly described in the question, which facilitates the transfer of learned comparison strategies across emotional dimensions. Notably, the DPO-CoT model achieves the strongest overall performance, with a larger performance gap compared to other variants than in the in-domain setting of Table~\ref{tab:main_results}. This result suggests that enhanced comparative reasoning through preference optimization is particularly robust for emotion transfer.

Finally, as illustrated in Fig.~\ref{fig:qualitative}, the DPO-CoT model provides interpretable explanations for its decisions even under unseen conditions, such as cross-language evaluation (English training and Chinese evaluation) and cross-emotion transfer (training on arousal and evaluating on dominance). These results indicate that the learned reasoning patterns generalize across domains and tasks, highlighting the potential of reasoning-guided preference learning for robust speech emotion comparison.


\begin{figure}[t]
  \centering
  
    \begin{promptbox}{Trained on MSP-Podcast (English), Evaluated on BIIC (Chinese)}
  \raggedright
    Clip 1 features a child’s clear, deliberate articulation with no hesitation, steady pacing, and confident vocal control, projecting assertive playfulness\textbf{. Clip 2 shows frequent fillers }\textbf{(``\begin{CJK*}{UTF8}{gbsn}\textbf{嗯，对，因为...}\end{CJK*}"),} self-corrections, and inconsistent pacing, reflecting uncertainty and lack of control. The child’s firm, unbroken delivery in Clip 1 contrasts sharply with the hesitant, fragmented speech in Clip 2, making Clip 1’s dominance more evident.
    \medskip \\
    \textbf{Answer: Clip 1}
  \end{promptbox}
  \begin{promptbox}{Trained on Arousal, Evaluated on Dominance}
    \raggedright

Clip 1 features a calm, measured delivery with minimal pitch or loudness variation, conveying a composed, professional tone. Clip 2 shows higher pitch variability and dynamic vocal effort, with expressive intonation and emotional emphasis on phrases like ``I'm so upset." \textbf{The increased vocal energy and emotional intensity in Clip 2 indicate greater dominance and control.}
\medskip \\
\textbf{Answer: Clip 2}
  \end{promptbox}
\vspace{-0.3cm}
  \caption{Qualitative examples of reasoning traces generated by the DPO-CoT model.}
  \label{fig:qualitative}
\end{figure}

\vspace{-3mm}
\section{Conclusions}

We studied whether LALMs can compare emotional content across speech rather than only predict emotions independently. By formulating ordinal speech emotion recognition as a pairwise decision task, we show that LALMs require task-specific adaptation but can perform reliable cross-audio comparison after SFT and DPO, achieving strong data efficiency relative to SSL-based preference models. Incorporating reasoning traces grounded in acoustic cues provides interpretable explanations for model decisions and improves robustness in cross-domain and cross-emotion settings. In particular, preference optimization with reasoning traces encourages the model to distinguish correct and incorrect reasoning patterns, reducing hallucinated explanations and improving comparative reasoning performance. Future work will explore extending this comparative reasoning framework to other speech attributes beyond emotion, including environmental, phonatory, prosodic, idiosyncratic, linguistic, and interpersonal dimensions.

\newpage
\section{Generative AI Use Disclosure}

Generative AI tools were used only to assist with minor language editing and stylistic polishing. The research design, experiments, analysis, and conclusions were entirely developed by the authors, who are fully responsible for the content of this manuscript.

\bibliographystyle{IEEEtran}
\bibliography{reference,mybib}

\end{document}